%% file: paper.tex
\documentclass[runningheads]{llncs}
\usepackage[utf8]{inputenc}
\usepackage[T1]{fontenc}

\input{builtins/std}

\usepackage{tikz}

\input{builtins/listings}
\usepackage[colorlinks=true, linkcolor=blue, citecolor=blue, urlcolor=blue]{hyperref}
\usepackage[capitalize,nameinlink]{cleveref}
\usepackage{subcaption}

\newcommand{\tool}{Cargo Scan\xspace}
\newcommand{\vet}{Cargo Vet\xspace}

\renewcommand{\orcidID}[1]{}

\begin{document}

\date{}

\title{Auditing Rust Crates Effectively}

\author{
Lydia Zoghbi\inst{1}\orcidID{0009-0008-5735-6768} \and
David Thien\inst{1}\orcidID{0009-0003-3840-051X}  \and
Ranjit Jhala\inst{1}\orcidID{0000-0002-1802-9421} \and
Deian Stefan\inst{1}\orcidID{0000-0002-7041-7464} \and
Caleb Stanford\inst{2}\orcidID{0000-0002-8428-7736}
}
\authorrunning{L. Zoghbi et al.}

\institute{
University of California, San Diego, USA \\
\email{\{lzoghbi,dthien,rjhala,dstefan\}@ucsd.edu} \and
University of California, Davis, USA \\
\email{cdstanford@ucdavis.edu}
}

\maketitle

\begin{abstract}
We introduce \tool{}, the first interactive program analysis tool designed to
help developers audit third-party Rust code.
Real systems written in Rust rely on thousands of transitive dependencies.
These dependencies are as dangerous in Rust as they are in other languages
(e.g., C or JavaScript)\dash---and auditing these dependencies today means
manually inspecting every line of code.
Unlike for most industrial languages, though, we can take advantage of Rust's
type and module system to minimize the amount of code that developers need to
inspect to the code that is potentially dangerous.
\tool models such potentially dangerous code as effects and performs a
side-effects analysis, tailored to Rust, to identify effects and track them
across crate and module boundaries.
In most cases (69.2\%) developers can inspect flagged effects and decide whether
the code is potentially dangerous locally.
In some cases, however, the safety of an effect depends on the calling
context---how a function is called, potentially by a crate the developer imports
later.
Hence, \tool tracks context-dependent information using a call-graph, and
collects audit results into composable and reusable audit files.
In this paper, we describe our experience auditing Rust crates with \tool{}.
In particular, we audit the popular client and server HTTP crate,
\texttt{hyper}, and all of its dependencies; our experience shows that \tool{}
can reduce the auditing burden of potentially dangerous code to a median of
0.2\% of lines of code when compared to auditing whole crates.
Looking at the Rust ecosystem more broadly, we find that \tool{} can automatically classify
$\sim$3.5K of the top 10K crates on \texttt{crates.io} as safe; of the crates that
\emph{do} require manual inspection, we find that most of the potentially
dangerous side-effects are concentrated in roughly 3\% of these crates.
\end{abstract}

\input{intro}
\input{motivation}
\input{tool}
\input{evaluation}
\input{discussion}
\input{rw}
\input{acks}

\bibliographystyle{splncs04}
\bibliography{ref}

\input{appendix}

\end{document}

%% file: builtins/std.tex
\usepackage{amsfonts,amssymb} %
\usepackage{comment}                         %
\usepackage{ifthen}                          %
\usepackage{minibox}                         %
\usepackage{multirow}                        %
\usepackage{suffix}                          %
\usepackage{graphicx}                        %

\usepackage{listings}
\usepackage{xcolor}
\usepackage{makecell}

\usepackage{array}
\usepackage{booktabs}
\usepackage{pifont}

\usepackage{xspace}

\newcommand\ie{i.e.,\xspace}
\newcommand\eg{e.g.,\xspace}

\newcommand{\rustinline}[1]{\lstinline[language=Rust]|#1|\xspace}

\newcommand{\para}[1]{\smallskip\noindent\textbf{{#1.}\xspace}}

\def\dash---{\kern.16667em---\penalty\exhyphenpenalty\hskip.16667em\relax}

  {\begin{list}{$\blacktriangleright$}%
    {\leftmargin=\parindent \itemsep=2pt \topsep=2pt
     \parsep=0pt \partopsep=0pt}}%
  {\end{list}}

%% file: builtins/listings.tex
\usepackage{listings}
\usepackage{xcolor}

\definecolor{tangoBackground}{HTML}{f8f8f8}
\definecolor{tangoDefault}{HTML}{000000}
\definecolor{tangoComment}{HTML}{8f5902}
\definecolor{tangoKeyword}{HTML}{204a87}
\definecolor{tangoString}{HTML}{4e9a06}
\definecolor{tangoNumber}{HTML}{0000cf}
\definecolor{tangoNameBuiltin}{HTML}{204a87}
\definecolor{tangoNameDecorator}{HTML}{5c35cc}

\lstdefinelanguage{Rust}{
    keywords={abstract, alignof, as, become, box, break, const, continue, crate, do, else, enum, extern, false, final, fn, for, if, impl, in, let, loop, macro, match, mod, move, mut, offsetof, override, priv, proc, pub, pure, ref, return, Self, self, sizeof, static, struct, super, trait, true, type, typeof, unsafe, unsized, use, virtual, where, while, yield, async, await, dyn, try},
    keywordstyle=\bfseries\color{tangoKeyword},
    ndkeywords={bool, char, f32, f64, i8, i16, i32, i64, i128, isize, u8, u16, u32, u64, u128, usize, str, String, Vec, Option, Result, Some, None, Ok, Err, Box, Rc, Arc, HashMap, HashSet, BTreeMap, BTreeSet, LinkedList, VecDeque, Cow, Cell, RefCell, Mutex, RwLock, PhantomData},
    ndkeywordstyle=\bfseries\color{tangoNameBuiltin},
    sensitive=true,
    comment=[l]{//},
    morecomment=[s]{/*}{*/},
    commentstyle=\itshape\color{tangoComment},
    stringstyle=\color{tangoString},
    morestring=[b]",
    morestring=[b]'
}

\lstdefinestyle{tangostyle}{
    basicstyle=\ttfamily\small,
    breaklines=true,
    captionpos=b,
    keepspaces=true,
    showspaces=false,
    showstringspaces=false,
    showtabs=false,
    tabsize=2,
    numberstyle=\fontsize{7}{9}\selectfont
}

\newcommand{\rustcode}[1]{\lstinline[language=Rust,basicstyle=\ttfamily\normalsize]|#1|}

\newcommand{\inputrust}[2][]{
    \lstinputlisting[
        language=Rust,
        style=tangostyle,
        xrightmargin=\parindent,
        basicstyle=\ttfamily\small,
        escapeinside={//},
        #1
    ]{#2}
}

%% file: intro.tex
\section{Introduction}

Rust is a compelling alternative to C and C++ for systems programming
as it provides memory safety by default through its strong static typing,
module-level encapsulation, array bound checks and compositional error handling.
However, modern software is predominantly \emph{other} people's code,
and arbitrary, third-party Rust can be just as dangerous as C or C++.
First, Rust provides an \rustinline{unsafe} keyword which allows developers
to bypass its usual safety guarantees, meaning that third party code must be
trusted to uphold Rust's complex memory safety invariants~\cite{evans2020rust,astrauskas2020programmers}.
Second, memory safety is not the only concern; memory \emph{safe} third-party
Rust code can invoke the operating system to read and write files, connect to
network endpoints, and execute arbitrary binaries---all of which
can potentially be dangerous.

\para{Problem: Vetting Third-Party Rust}
Currently, Rust teams \emph{audit} third-party code (crates)
to mitigate the risk of supply chain attacks.
This is a manual, painstaking process as real-world systems
depend (transitively) on hundreds to thousands of crates.
Efforts like Mozilla's \vet~\cite{cargo-vet} streamline auditing by allowing
Rust teams to record their results and sign off on a crate's safety, enabling
others to \emph{trust} these results and lower their workload.
In practice, however, this trust-based approach does not really
simplify the review process (\S\ref{sec:overview}).
Further, as \vet's audit results are just approvals with unstructured
comments, even audited crates can require a look over, since a crate's
publicly exposed function might be safe only in specific contexts.
Finally, identifying what code needs auditing---even within an
\rustinline{unsafe} block~\cite{rustlangExtendingUnsafe}---and
then determining its actual safety remains challenging.

\para{Auditing Goals}
We identify two key goals for a tool-assisted auditing process.
\begin{enumerate}
\item \emph{Audit the code that matters.}
Developers should not have to inspect every line of code; they should
only audit code that is indeed potentially dangerous.
Most widely-used languages make this near impossible (e.g., C/C++'s memory
unsafety, Java's reflections and class loading, and most of JavaScript's
features make it hard to look at a piece of code and be confident it's safe);
Our goal is to leverage Rust's type and module system to get
global safety properties by auditing only a small fraction
of the code.
\item \emph{Track context-dependent information.}
Developers should not have to audit already-audited code.
Currently, they must as some functions are only safe
\emph{sometimes}, i.e., depending upon the calling context.
Instead, audits should be \emph{composable}
and track such information in the audit report
itself; to inform developers when they use such
potentially dangerous functions, thereby propagating
the context-sensitivity to any callers of this function.
\end{enumerate}

\para{Auditing Rust Effectively}
We built \tool{}, an interactive Rust auditing tool, with these goals in mind.
\tool{} is founded on central insight borrowed from
functional programming: we can view potentially
dangerous operations as \emph{effects}, which are
flexible enough to model a wide variety of potentially
dangerous operations, from low-level memory operations
(e.g., raw pointer dereference) to high-level operating
system calls (e.g., file reads and writes).
However, rather than tracking effects in the
type system (e.g., as in Haskell), we track them via
an off-the-shelf static analysis that lets
developers focus on the code that matters
while tracking and propagating effects
context-sensitively.
In this paper, we present the design and implementation of \tool and report on
our experience auditing crates with \tool.
Our contributions are threefold.

\begin{figure}[t]
    \includegraphics[width=10cm, keepaspectratio]{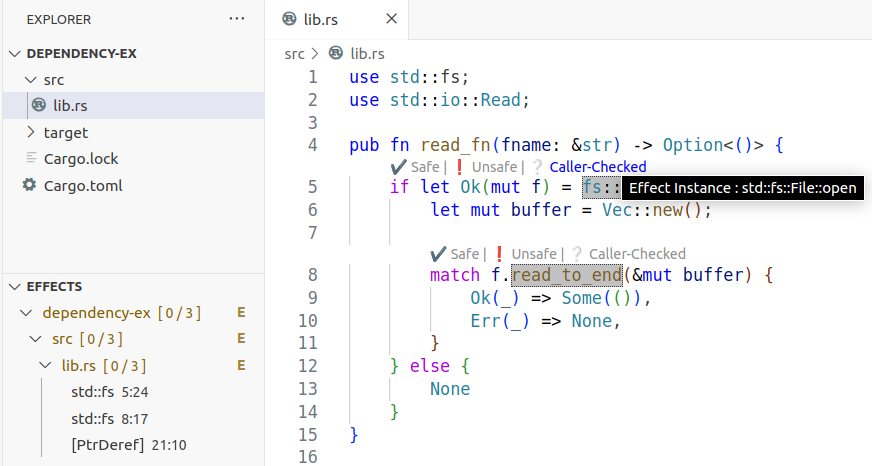}
    \caption{Interactive auditing with \tool{} in VSCode.}
    \centering
    \label{fig:ext_overview}
\end{figure}

\para{1. Effect Analysis}
Our first contribution addresses the first auditing goal via a static
analysis pass that identifies \emph{potential side effects}
in the source code (\S\ref{sec:tool:analysis}).
Our effect model (\S\ref{sec:tool:model}) and analysis builds on the long line of work on effects
analysis, but crucially exploits Rust's type system and our problem domain
to make auditing practical.
First, we observe that the borrow checker in Rust,
combined with module-level abstraction and the explicit safe/unsafe
scoping, already makes mutation and aliasing tractable,
contrary to other languages.
This allows us to focus on a \emph{function-level} analysis
(by associating side effects with each function call)
rather than a traditional \emph{data-level} analysis
that is necessary in languages like Java
~\cite{JoanAudit,livshits2005,andromeda,findbugs,jpure,rountev2004precise,le2005jit,milanova05parameterized,purity-analysis,YangHHK15,tao2012escape}.
Moreover, this analysis lets us ensure that code that does not have potentially
dangerous side effects is not presented to the user at all.
Second, we observe that complex language features---closures, function
pointers, traits and generics---which make analysis hard in the general
case don't need to make auditing hard(er).
Our insight is to ``lift'' these features into the effect model and ask
developers to audit, for example, code creating closures much like
they would audit code opening files.

\para{2. Interactive Auditing}
Our second contribution addresses the second goal via an interactive auditing
process that prompts the user with potentially dangerous effect sites (\S\ref{sec:tool}).
The user then marks each site with an annotation that indicates it's (un)safe
or that its safety is context-sensitive.
\tool{} captures this information in the audit file, hence, unlike \vet, developers
don't need to trust a library in whole, nor reaudit it: they can read the
audit report (using \tool{}) and just inspect the calling context of
context-dependent functions.

We implement the \tool interactive process as a VSCode extension.
This lets developers easily assess the safety of any particular effect and annotate
it inline: they can inspect the surrounding code and take advantage of integrations
like \texttt{rust-analyzer} to inspect types, definitions, and otherwise navigate the
call graph (and codebase).
Our experience auditing the \texttt{hyper} networking library and all
of its dependencies highlights that this is especially useful for complex codebases
and codebases where many effects are context-sensitive (and navigating between call
sites and definitions is crucial).
Anecdotally, this extends to commercial use cases;
the authors have experimented with \tool{} for over a year in production
to audit Rust crates for a security critical product,
and have found that it reduces the time to complete code reviews for new crates in
pull requests from multiple days to a few hours.

\para{3. Evaluation}
Our third contribution is an evaluation of \tool{}
which shows that tool-assisted auditing
reduces the auditing burden and identifies cases of
context-sensitive safety (\S\ref{sec:evaluation}).
We first show that out of the top 10K most used crates, 3434 of them
are \emph{pure}, \ie have no dangerous side effects at all,
meaning they can be declared safe to use \emph{sans}
any human auditing, and further, that 85\% of side
effects are contained in just 3\% of crates.
This means \tool{} can vet most crates with little to no auditor effort.
We then report on our experience auditing the
\texttt{hyper} networking library and all of its
dependencies.
Our audit shows that 30.8\% of functions' safety is context-sensitive, and
5.2\% of functions' safety is context-sensitive across package boundaries; this
means auditing code in practice crucially relies on tracking effects across
functions (and packages).
Overall, though, this does not impose a huge burden: we find that
tool-assisted auditing still only requires auditing a median of 0.2\% of lines
of code (relative to auditing the code without \tool{}).
Finally, if the top 10K crates are representative of the broader Rust ecosystem,
this means we can use \tool{} to raise the security bar of Rust as a
whole with a modest manual effort.

%% file: motivation.tex
\section{Motivation}
\label{sec:overview}
\label{sec:motivation}

Organizations like Mozilla, Fastly, and Google manually audit third-party code
before using it in security critical systems like Firefox, Wasmtime, and Chrome.
Until recently, much of this effort was duplicated across
organizations, in the absence of a common platform to share results, and
simply trusting a library because one organization depends on it doesn't
provide any inherent guarantees.

Mozilla built \vet to tackle precisely this problem~\cite{cargo-vet}.  With
\vet, developers audit (or vet) packages against shared criteria
(e.g., ``safe to run'' locally or ``safe to deploy'' in production).
The tool then records their result in a common format, alongside the package
version (or the delta between versions) they audited and any (unstructured)
notes.
This allows organizations to share the audit burden; a team can skip auditing
packages by \emph{importing} audits of a trusted entity
(\eg Chrome's audit records), or by marking authors as trusted---especially
if say Mozilla and Google trust them.

\subsection{Manual Auditing Does Not Scale}

In practice, importing external audits and marking prolific developers as
trusted, still leaves us with a lot of code to audit.
Here, \vet doesn't help: auditing even strongly-typed Rust code is a
manual, painstaking, and error-prone process.

\para{1: We need to audit everything or risk missing dangerous code}
Consider auditing the \texttt{hyper} HTTP client and server library, one of the
most popular libraries on \texttt{crates.io} with over 406 million downloads.
The library is core to popular web frameworks like \texttt{axum}, which use
``\rustcode{#![forbid(unsafe_code)]} to ensure everything is implemented
in 100\% safe Rust.~\cite{axum}''
Unfortunately, this notion of safety is not that meaningful in practice;
\texttt{hyper} and its transitive dependencies \emph{do} have
\rustinline{unsafe} code.
Moreover, code performing potentially dangerous operations (e.g.,
complex code that accesses the filesystem, network, etc.)---although
considered safe by Rust's standards---is still present in these packages.
We thus need to vet \texttt{hyper} and its dependencies are ``safe to
deploy''---and today that would mean manually sieving through 160K lines of
code that spans the 30 packages.
From our and other developers' experience auditing crates, finding
potentially dangerous code---beyond the obvious \rustinline{unsafe}
blocks---using off-the-shelf tools like grep is hard.
In practice, this doesn't mean sifting through 160K lines of code, it
means giving up on the audit, or accidentally missing potentially dangerous code.
This is particularly unfortunate because---as we show in
\S\ref{sec:common_effects}---large fractions of many packages
\emph{can} be safely vetted completely automatically.

\para{2: We need to reason about calling contexts}
Within \texttt{hyper} itself, there are lots of calls into the standard
library---to access the filesystem and network---but unlike
\rustinline{unsafe}, developers don't have to explicitly annotate this code as
potentially dangerous.
For example, this code from \texttt{hyper}, \emph{looks} safe:
\inputrust[frame=lines, basicstyle=\ttfamily\scriptsize]{code/motivation_examples/bind.rs}
\noindent The function is just constructing a new \rustinline{AddrIncoming} and is easy to
skip---both on manual inspection and with tools like (sem)grep.
However, it is only by inspecting \rustinline{AddrIncoming::new} that we see that
it, and in turn, \rustinline{bind}, are potentially
dangerous---the call binds and listens to the socket address.
For this same reason, we would need to inspect the calling context where
\rustinline{bind} is used---and potentially the calling context where that
function is used.

\para{3: We need to reason about complex types}
Higher-order functions and polymorphic code makes things harder.
Rust's traits in particular, which are similar to interfaces and abstract classes in other languages,
make it hard to reason about and analyze (polymorphic) code.
Consider the \rustinline{Subscriber} trait definition from the
\texttt{tracing\_core} package, which provides the core functionalities for
application-level tracing to packages like \texttt{hyper}:
\inputrust[frame=lines, basicstyle=\ttfamily\scriptsize]{code/motivation_examples/downcast_raw.rs}

\noindent
The default implementation of \rustcode{downcast_raw} invoked in \rustcode{downcast_ref}
is safe, as it ensures to only return valid pointers, and is easy to audit.
The danger lies when other implementations of this trait don't satisfy the safety requirements.
Hence, \rustcode{downcast_ref} is not actually safe in all
contexts---and auditing this trait alone is insufficient; we need to audit all its
implementations too.

\para{4: We need to track effects across packages}
Reasoning about potentially dangerous code within a single crate is hard and
non-local.
Consider CVE-2021-45712~\cite{cve-2021-45712} in the \texttt{rust-embed}
project.
\texttt{rust-embed} is a Rust custom derive macro---a type of procedural
macro that automates the implementation of traits for structs, enums, or unions.
It takes a user-defined directory and either (1)~embeds files into the binary
in release mode or (2)~reads files directly from the file system in debug mode.
By design, the \texttt{rust-embed} crate and its related packages (try to)
enforce filesystem isolation, i.e., the derive macro is constrained to only read
files in the user-defined directory.
In debug mode, however, the implementation did not properly
sanitize input paths---and broke this isolation guarantee (as
RUSTSEC-2021-0126~\cite{rustsec-2021-0126} and
CVE-2021-45712~\cite{cve-2021-45712} describe in detail).

This was easy to miss in an audit given many intertwined crates were involved.
The \texttt{rust-embed} crate defines the \rustcode{RustEmbed} trait
for the procedural macro, implemented in the \texttt{rust-embed-impl} crate,
which calls into the function \rustcode{read_file_from_fs} from
\texttt{rust-embed-util} to access the file system.
In debug mode, insufficient filepath sanitization in the buggy trait
implementation allows unrestricted access to files beyond the configured
directory.
While \rustcode{read_file_from_fs} could have enforced access checks
itself, instead of transfering this responsibility to its callers, there is no
mechanism for reviewers to indicate this while auditing.
This function is not fundamentally dangerous; its safety is context-sensitive and
depends on how it is invoked.

Spotting such bugs during auditing requires reasoning across multiple crates,
trait implementations, and runtime configurations, and existing tools provide
little assistance: (1) simple pattern searches (\ie with \texttt{grep})
fail as there is no \rustcode{unsafe} code or effectful standard library
call involved in the bug's location, and (2) tools like \vet focus on auditing
packages in isolation---and fundamentally this breaks down on many real-world
packages.

\subsection{Our Solution}

\begin{figure}[t]
    \includegraphics[width=11cm, keepaspectratio]{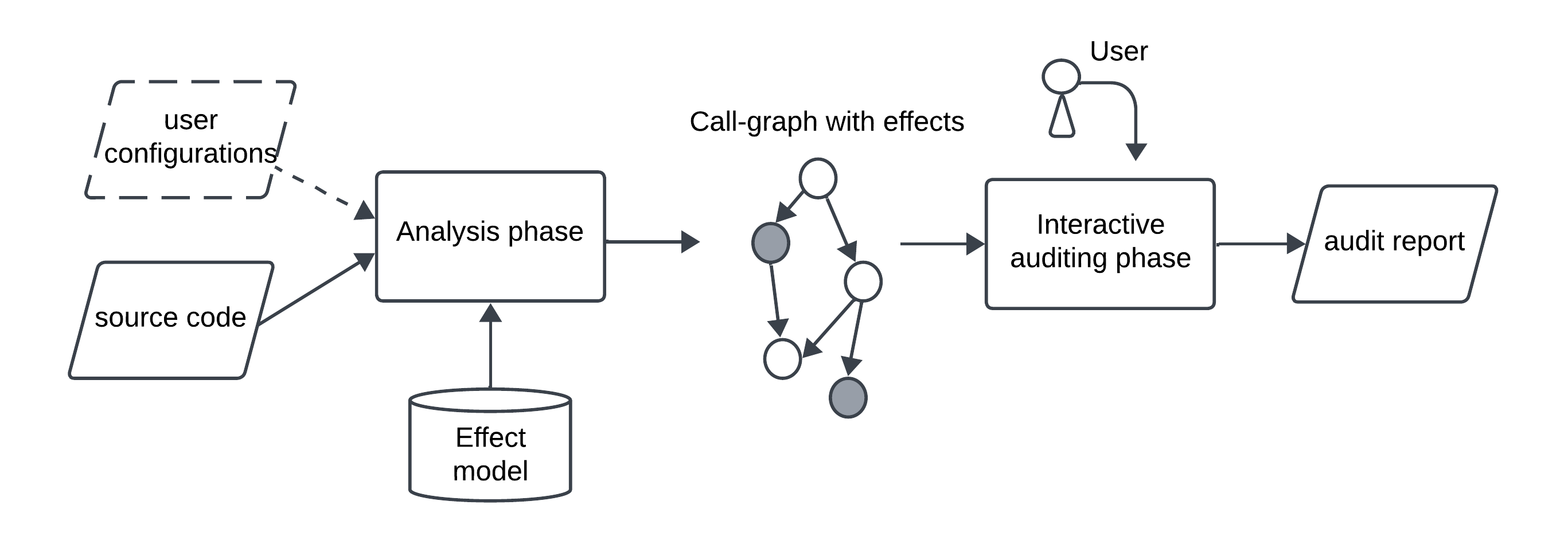}
    \caption{\tool{} overview.}
    \centering
    \label{fig:overview}
\end{figure}

Thus, while \vet provides clear benefit to users,
it doesn't help auditors and developers in the process
of actually reviewing code.
We developed \tool{} to assist in the lengthy and tedious auditing process. Our
goal is to give developers an intermediate step in the space between purely
manual inspection during auditing and static analysis-based bug-finding tools
used by other ecosystems beyond Rust.  Our semi-automated architecture is
summarized in \Cref{fig:overview}.

%% file: tool.tex
\section{\tool{} Overview}
\label{sec:tool}

\tool{} addresses the challenges laid out in \S\ref{sec:overview}
by pinpointing (only) the
locations where potentially dangerous code is executed,
and guiding users through the function call-stack to make
sure that the code is only used in safe contexts.

\subsection{Effect Model}\label{sec:tool:model}
An \emph{effect} is any function behavior
beyond what appears in its signature (input/output types),
occurring during execution or afterwards.
As Rust permits mutation via explicit mutable reference types
like \rustcode{&mut T} and mutable cells like
\rustcode{Cell<T>}, we do not consider mutation
itself as an effect by default.
Although interior mutability could be considered an effect,
mutation inside interior mutable abstractions, like
\rustcode{Cell}, \rustcode{UnsafeCell}, and \rustcode{Mutex},
cannot alone lead to memory unsafety or other system side effects.
Such abstractions ultimately get flagged when using unsafe trait functions
or other potentially dangerous operations.

Our effect model consists of three categories of effects, as summarized in \Cref{tab:effects}:
(i) \textit{unsafe effects}, (ii) \textit{system effects}, and (iii) \textit{higher-order effects}.
We conservatively overapproximate the sink patterns for our analysis
by comprehensively inspecting the standard library's modules to find
which of them invoke syscalls.
These patterns are (in effect) extended further whenever context-sensitive
effects from dependencies are propagated to the target package.
Additionally, modelling closures and function pointers as effects at creation time
makes auditing easier; it's easier to audit a closure passed to higher-order
functions like \texttt{map} than it is to audit every application of it.
Indeed, in our experience this design choice greatly simplifies auditing, as, in Rust,
many closures are purely functional code.
Modelling them as effects at use time would require auditing all locations that
perform a map or filter over a generic function, including marking \texttt{map}
and \texttt{filter} in the standard library as effectful, leading to many false
positives.
In sum, the effect model is sufficient for the fundamental types of
effects that can break Rust's safety guarantees and introduce security
risks, like memory manipulation, system interactions, or communication
with other languages.
We describe in detail these above categories in the Appendix~\ref{sec:appendix}.

\begin{table}[t]
  \centering
  \caption{\tool's effect model.}
  \begin{tabular}{>{\centering\arraybackslash}p{9em} >{\centering\arraybackslash}p{27em}}
      \toprule
      \multicolumn{2}{c}{\textbf{Effect Model}} \\
      \midrule
      \textit{Unsafe Effects}
      & FFICall, FFIDecl, StaticExt, StaticMut, UnsafeCall, UnionField, RawPointer \\
      \midrule

      \textit{System Effects} \newline (SinkCall)
      & std::fs, std::io, std::os, std::ffi, std::net, std::env, std::arch, std::path, std::mem, std::simd, std::panic, std::process, std::backtrace, std::intrinsics, libc, winapi \\
      \midrule

      \textit{Higher-order Effects}
      & FnPtrCreation, ClosureCreation \\
      \bottomrule
  \end{tabular}
\label{tab:effects}
\end{table}

\subsection{Effect Analysis}\label{sec:tool:analysis}
Our analysis operates directly on a \emph{target package}'s
source code in a single, mostly syntactic pass.
\tool{} parses the input package into an AST and performs a lightweight,
pattern-matching traversal to identify effects.
For example, each call or expression node is compared against a predefined
set of canonical library paths (\eg \texttt{std::fs}) or operation categories
(\eg \texttt{RawPointer}) in our effect model.
Performing our analysis on AST instead of a lower representation isn't fundamental,
but offers some useful advantages.
It facilitates the direct mapping of effects back to source code and the precise
calculation of audited lines of code before desugaring.
Most effects can be sufficiently identified via pattern-matching, but
when necessary, \tool{} leverages semantic information from \texttt{rust-analyzer}~\cite{rust-analyzer},
which operates on both HIR and MIR.
This reduces false positives by resolving types, names, and disambiguating constructs
such as raw pointer dereferences or identically named functions and trait methods.

\tool{} also builds a call graph for the input package and associates each
node with a set of effects inline.
We implement a variation of taint and effect analysis, where the identified
effects can be considered as the tainted sources and the publicly exposed
functions of a crate are the sink locations.
Rust's type system makes mutation and aliasing tractable, and thus,
allows us to perform a scalable and lightweight intraprocedural analysis
without pointer, alias, and data-flow analyses, and capture a wider variety
of dangerous behaviors.

\subsection{Safety Annotations}\label{sec:tool:annotations}
\label{sec:tool_annotations}

\tool{} provides users with three safety annotations to mark effects
during an audit and indicate their safety: (i)~\emph{safe},
(ii)~\emph{unsafe}, or (iii)~\emph{caller-checked}.

\para{Safe/Unsafe Annotation}
The first two annotations -- \emph{safe} and \emph{unsafe} -- are applicable
when an effect's safety can be determined \emph{locally} i.e.,
from the surrounding code.
In the snippet below, the \rustcode{get_unchecked} effect is safe,
because the vector's length is checked before accessing it.
\inputrust{code/tool_examples/safe_annotation.rs}

\para{Caller-Checked Annotation}
The third annotation -- \emph{caller-checked} -- is applicable when
the code's safety (or lack thereof) cannot be established in isolation
and depends on the \emph{calling context}.
For example, the function
\inputrust{code/tool_examples/cc.rs}
\noindent may be called innocuously to read a configuration file,
or mischievously to read a user's private keys, depending on the
value of the parameter \texttt{f}.
Such effects are annotated as \emph{caller-checked} and \tool{} utilizes
the call-graph from the analysis phase to propagate them to all
callers of the function (here, \rustcode{read_file}).

\subsection{Interactive Auditing}\label{sec:tool:interactive}

\tool{} facilitates a systematic vetting of the identified effects by
allowing the user to interactively audit the code at effectful
locations.
We initially implemented \tool{} as a command-line interface similar to
interactive staging for \texttt{git add} for the interactive auditing.
After using that interface to audit numerous crates, we realized that in
practice navigating the source code and effects, as well as tracking our
progress during auditing was too restrictive.
Thus, we implemented a VSCode extension (\Cref{fig:ext_overview}) to
tackle these limitations and leverage the interactive features of an
editor for more flexibility.

\subsection{Cross-Package Audits}\label{sec:tool:multi}
\label{sec:tool_audit_files}

To fully audit a (client) package, all its dependencies must
first be audited to see which public methods are
caller-checked (which would then necessitate auditing
their call-sites in the client package).
As shown in the \texttt{rust-embed} example in \S\ref{sec:motivation},
vulnerabilities that originate from caller-checked public functions in other
packages happen in practice.
\tool{} tracks public caller-checked functions from dependencies
that are transitively reachable from the top-level package.

\para{Default Audits}
\label{sec:tool_default_audits}
Auditing all of a project's dependencies in full produces the most precise
accounting of possible effects, but requires significant effort at scale.
\tool{} automates this process by generating \emph{default} audits---sound,
conservative approximations of manual reviews---by marking every effect as
caller-checked and propagating them up the stack.
Treating all effects as context-sensitive computes every potential propagation path
and captures the upper bound of the code requiring review for all possibly dangerous
operations.
Consequently, users can restrict their attention in auditing only those public
caller-checked functions from dependencies called in their application.
This allows \tool{} to isolate only relevant code for auditing across dependencies,
similarly to how we identify effectful locations within a single package.
Crates that yield overly conservative default audits can be audited first,
and any library calls deemed safe are then removed from dependent packages'
effect sets.

%% file: evaluation.tex
\section{Empirical Evaluation}
\label{sec:evaluation}

We evaluate \tool{} by asking three questions:
\begin{itemize}
  \item How common are effects in popular packages (\S\ref{sec:common_effects})?
  \item Is context-sensitive effect propagation valuable (\S\ref{sec:eval_cc})?
  \item Does \tool{} make auditing more efficient (\S\ref{sec:efficient})?
\end{itemize}

\para{Experimental setup}
We ran the experiments of \S\ref{sec:common_effects} on a machine with AMD
Ryzen ThreadRipper 3960X with 24 cores (and 48 threads), 128GB RAM, with
Ubuntu 24.04 and Rust 1.90.
We used custom scripts with 16 parallel threads (due to memory restrictions).
We performed the audits of sections \S\ref{sec:eval_cc} and \S\ref{sec:efficient}
both on the previous machine and on another with Intel i7-8650U, 16GB RAM, with
Ubuntu 24.04 and Rust 1.90.
All experiments were run as of October, 2025.

\subsection{How Common are Effects?}
\label{sec:common_effects}

We first look at how common effects are in Rust
by using \tool{} to analyze the effects found
in the top 10K most downloaded packages on
\texttt{crates.io}, the central public Rust crate
repository.
For our ecosystem analysis, we analyzed each crate individually,
without tracking effects across dependencies.

\begin{figure}[t]
    \centering
    \includegraphics[width=0.5\textwidth]{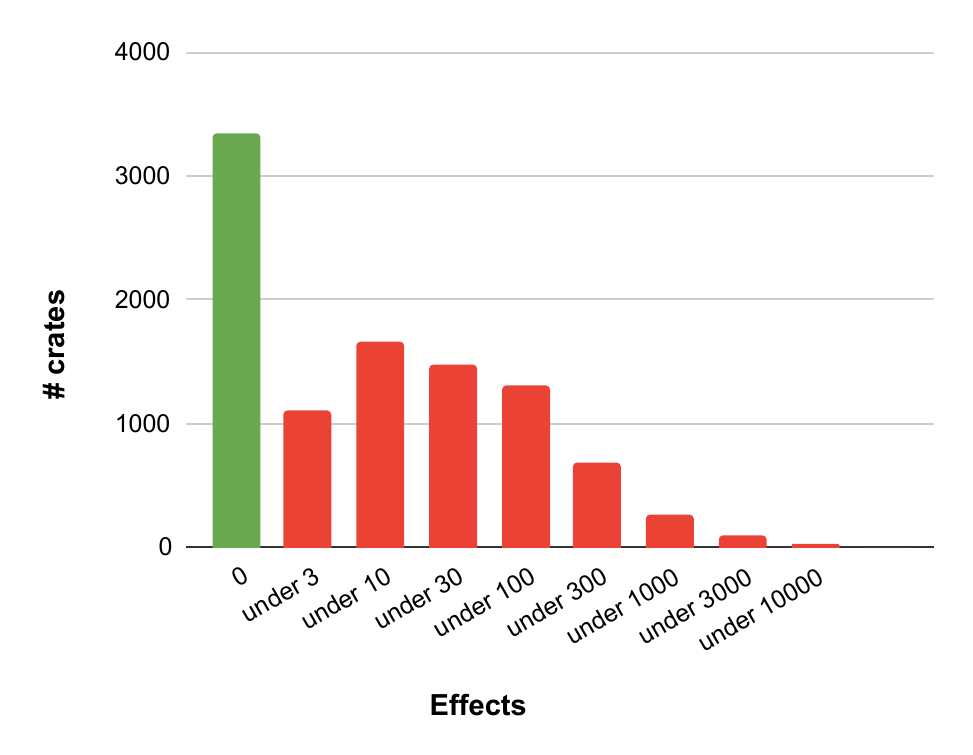}
    \caption{Distribution of top 10K crates by number of effects.}
    \label{fig:ecosystem-1-effects-per-crate}
\end{figure}

\para{Concentration of effects}
\Cref{fig:ecosystem-1-effects-per-crate} shows the distribution of crates by total number of effects,
divided by order of magnitude.
We find that a large number of crates are
entirely or mostly \emph{pure}, \ie effect-free:
in particular, slightly over a third of crates (3434) contain no instances of
side effects, and over a quarter (2778) contain under 10 effect
instances.
Further, most effects (approx. 85\%) are concentrated in a small number
of crates (approx. 3\%).

\begin{figure}[t]
  \centering
  \begin{subfigure}[t]{0.35\textwidth}
  \includegraphics[width=\textwidth]{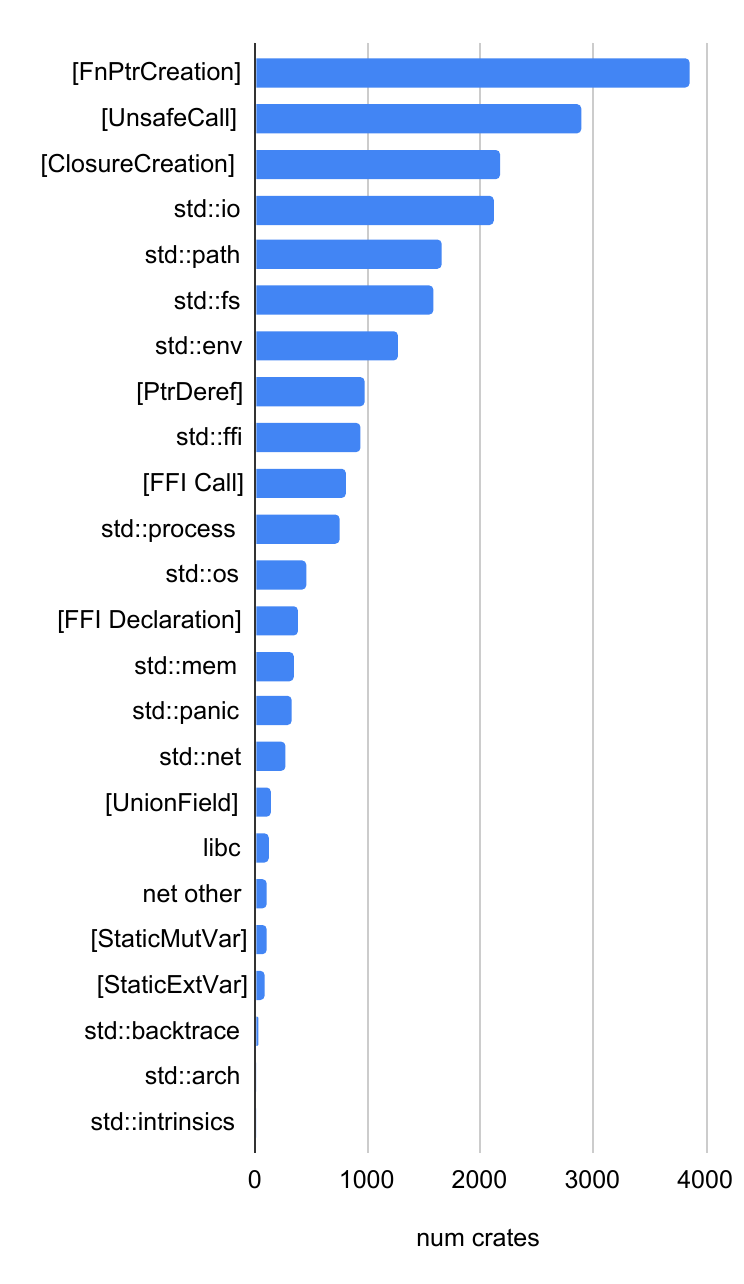}
  \caption{
    Number of crates with at least one occurrence of the effect in the
    top 10K crates.
  }
  \label{fig:ecosystem-2-crates}
  \end{subfigure}
  \qquad
  \begin{subfigure}[t]{0.55\textwidth}
  \includegraphics[width=\textwidth]{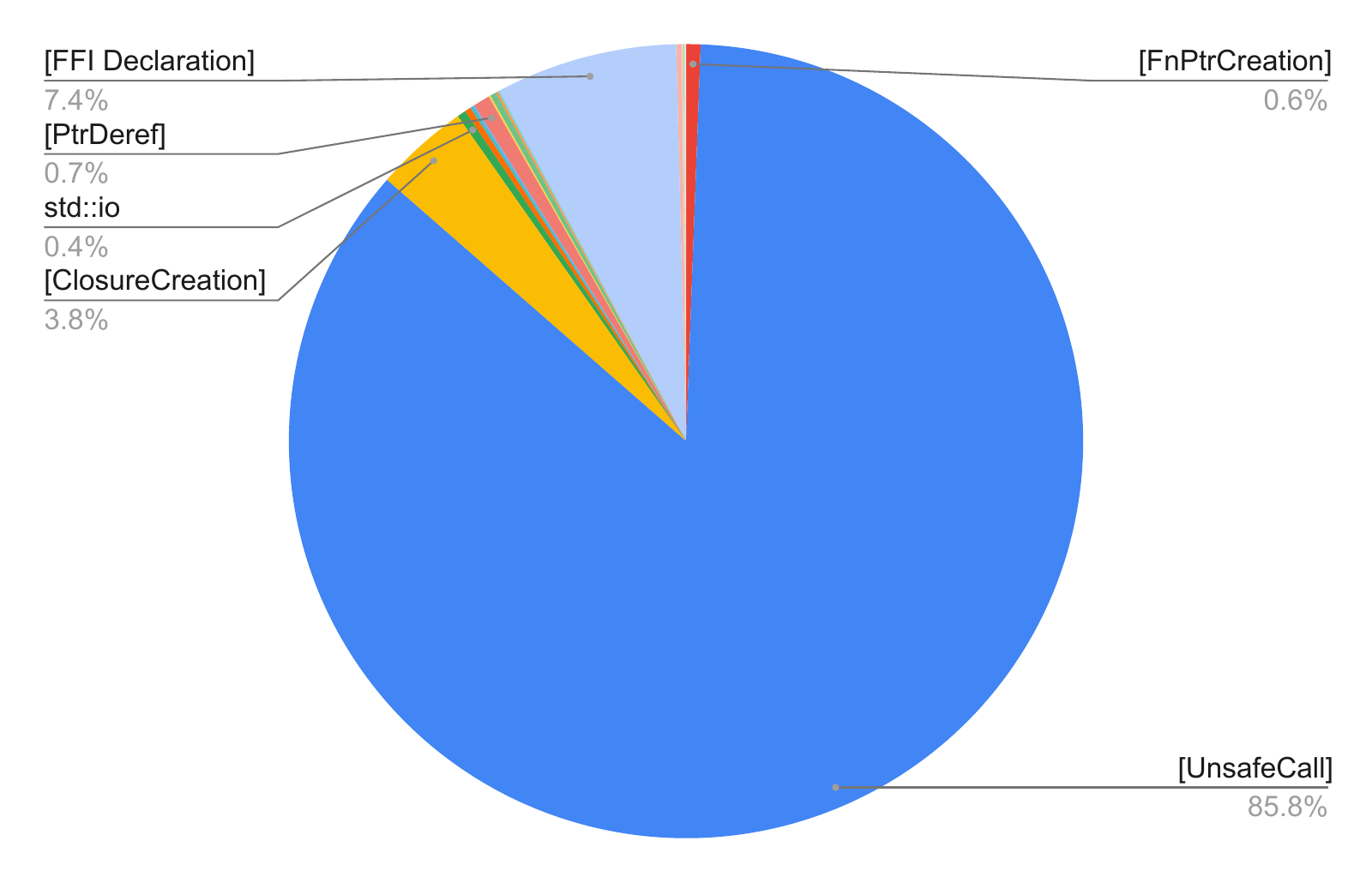}
  \caption{
    Relative frequency of effects by total instances
    across all top 10K crates.
  }
  \label{fig:ecosystem-2-pie-chart}
  \end{subfigure}

  \caption{Distribution of side effects by type.}
  \centering
\end{figure}

\para{Frequency of effect types}
\Cref{fig:ecosystem-2-crates} shows the frequency with which different effects
occur in the most popular packages.
We use the effect categories described in \S\ref{sec:tool:model}, and further
distinguish \texttt{SinkCall}s by the different libraries the calls originate
from.
In total, we include 24 effect patterns: Rust unsafety (pointer dereferences,
unsafe function calls, FFI, etc.), standard library effects (IO, file system,
network, and syscalls), and higher-order effects (creation of a closure or
function pointer which may be effectful).
The effects which affect the greatest number of \emph{crates} are higher-order
effects (function pointers and closures), unsafe function calls, standard I/O operations
(sinks), and raw pointer dereferences as shown in \Cref{fig:ecosystem-2-crates}.
The results are slightly different when broken down by number of
\emph{instances} of each effect across all top 10K crates, as demonstrated in
\Cref{fig:ecosystem-2-pie-chart}; in this case FFI function declarations
rise to the top, due to a small number of crates which consist mainly
of FFI bindings.
To gain a deeper insight, we also filtered out crates particularly dedicated for
FFI bindings, commonly known as \textit{sys-} crates.
The total number of crates that contain an FFI declaration effect drops from 379 to 190,
and the total frequency of this effect type across the top 10K crates is now 1.7\% compared to the initial 7.4\%.

\subsection{Are Context-Sensitive Effects Valuable?}
\label{sec:eval_cc}

Context-sensitive effects (tracked via the \emph{caller-checked} annotation \S\ref{sec:tool_annotations})
give auditors a more precise model of program safety,
but propagating effects to their call-sites may require more auditing effort.
To evaluate this we ask:
\begin{itemize}
\item \textbf{Q1}: What percentage of effects are caller-checked?
\item \textbf{Q2}: What is the average call-stack we audit for caller-checked effects?
\item \textbf{Q3}: How many caller-checked effects reach a crate's public API?
\end{itemize}
For this experiment we audited \texttt{hyper},
a widely used HTTP library, and all its dependencies in its default build in a
topologically sorted order (seven crates in total), and counted the effects that
require context-sensitive reasoning.
To answer \textbf{Q1}, among the packages we audited, the average percentage
of caller-checked effects was 30.8\%.
Regarding \textbf{Q2}, we had to inspect and annotate an arithmetic mean of
3.1 locations per effect including its origin.
Overall, this means most effects are only propagated to a small number of
locations, i.e., potentially dangerous side effects are well-encapsulated,
and reasoning about them requires only modest additional effort.
Finally, for \textbf{Q3}, fully audited packages (\ie for which we audited
all effects) had on average 5.2\% of their public functions marked as caller-checked.
Although caller-checked effects across the crate boundary are less common than
caller-checked effects within a crate, this finding still suggests the importance of
making audits composable by surfacing such effects in reports.

\begin{figure}[t]
  \centering
  \includegraphics[width=0.8\textwidth]{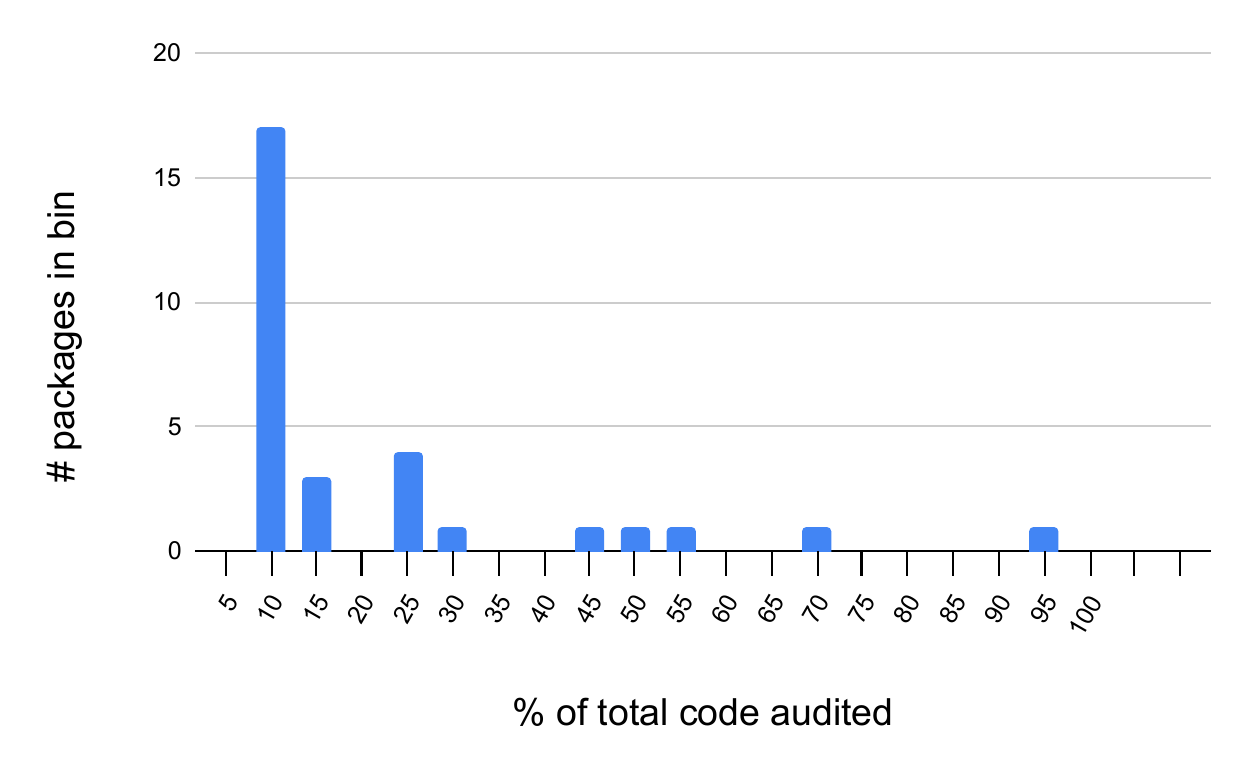}
    \caption{Probability density function for the percentage of code audited.
    The bars correspond to the number of packages whose percentage code audited
    is in the range $(x-5, x]$ (e.g., in the range $(0\%, 5\%]$).}
  \label{fig:audited-cdf}
\end{figure}

\subsection{Does \tool{} make Auditing more Efficient?}
\label{sec:efficient}

\begin{figure}[t]
  \centering
    \includegraphics[width=\textwidth]{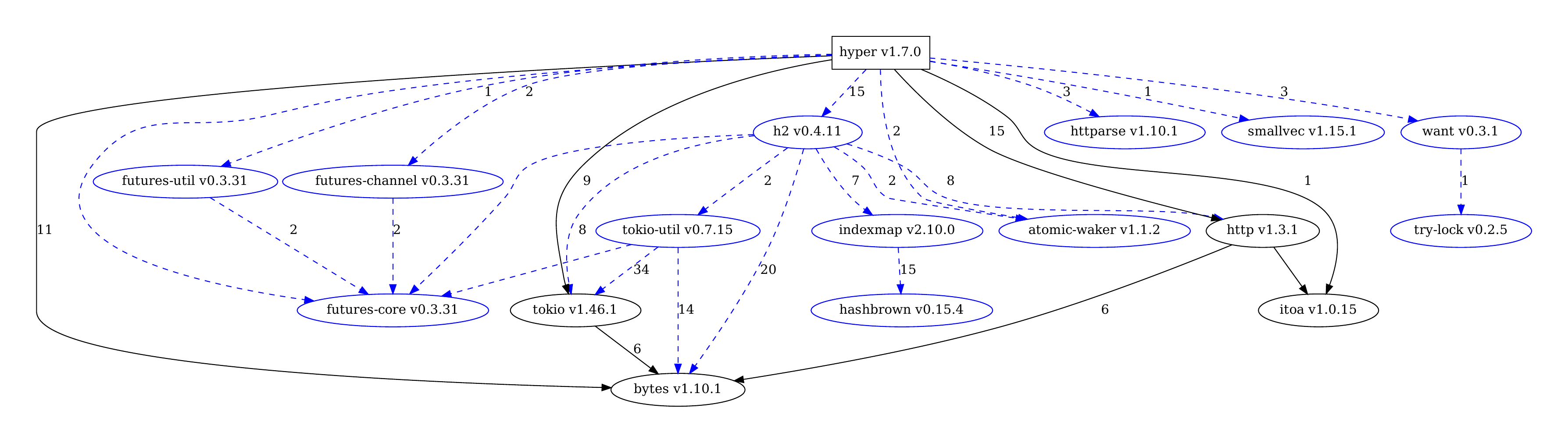}
  \caption{Subset of the dependencies with effects that flow into \texttt{hyper}.
  Edge weights represent the number of public caller-checked (CC) functions.}
  \label{fig:effect_propagation}
\end{figure}

As a proxy for measuring \tool{}'s use in practice, we instead measure auditing
efficiency in terms of the total amount of audited code, which directly impacts
auditing effort and time.
To evaluate \tool{} we use the same case study as in the experiment of section
\S\ref{sec:eval_cc}, but instead of using the default build we enable \emph{all} optional features to account
for all cases, and generate default audits for all dependencies.
This overapproximates all their potential effects and their propagation
in parent packages, i.e., caller-checked effects in calls \emph{across}
package boundaries (\S\ref{sec:tool_default_audits}).
Although default audits conservatively result in more audit locations,
\tool{} only presents those that are reachable from the root crate,
in this case \texttt{hyper}, so we focus our auditing effort to a small
subset of the package set.
Specifically, from a total of 30 packages for \texttt{hyper} and its normal and
optional dependencies, \tool{} only presents 17 of them for auditing, as shown in
\Cref{fig:effect_propagation}.
The edge weights represent public functions marked as caller-checked in dependent crates.

\para{Auditing approach}
We audited the crates in their topological order, i.e., we audit a crate's dependencies
before the crate itself.
In this case, the bottom crates of the dependency graph contain the highest percentage
of effects and are ``reused'' by other packages.
When effects could not be resolved locally in a crate, we reviewed call-sites
in parent crates, recursively, until we could determine their safety.
\tool{} automatically updates the set of reachable effects across dependencies and
eliminates any corresponding audit locations during auditing.
This significantly minimizes the total amount of audited code across
the dependency graph, especially when it includes large libraries (e.g.,
\texttt{tokio}) of which only a small fraction of the API is called.

\para{Quantifying auditing effort}
We determine auditing effort by measuring how much code we need to
review as a percentage of the \emph{total code} of the package as
analyzed by \tool{}.
Measuring ``reviewed code'' can be tricky as auditing an effect
requires understanding its surrounding context---including
things like the parameters of a function and how the result of an
effect is used.
Hence, rather than mark individual lines as relevant, we approximate ``reviewed
code'' by assuming we have to inspect each entire function an effect appears in.
To quantify the total amount of audited code in our case study,
we collect all functions with effects in \texttt{hyper} itself.
Since we use default audits as a starting point for each dependency, though,
we handle them differently than the top-level crate:
we track the effects ultimately marked as \textit{safe}
or \textit{unsafe} during our review---indicating we manually
audited their containing functions---along with all
caller-checked effects still reachable from the root.

\para{Distribution of audited effects}
We find that \tool{} significantly reduces the amount
of code an auditor has to look at, and that most of
the auditing effort is concentrated in a few crates
with many effects.
The arithmetic mean percentage lines of code we audit
per package is 13.1\%, with a median of 0.4\%.
The full distribution is in \Cref{fig:audited-cdf}.
We see a skew here consistent with what we expect from
\S\ref{sec:common_effects}.
Specifically, the packages with a high percentage of audited
code are primarily packages which are more commonly reused.

\para{Cross-package audits}
Counterintuitively, we find that using \tool{}'s default audits actually
reduces the total amount of code we need to audit; we audited on average
12.4\% of lines of code.
In contrast, in the experiment of section \S\ref{sec:eval_cc}, we
audited an average of 28.1\% for the same packages (\Cref{tab:audits_comparison}).
From this, we conclude that \tool{}'s default audit generation improves the
auditing workflow and reduces the total amount of code that needs
inspection across dependencies.

\para{Threats to validity}
Auditing is still a manual process and results may be impacted by various
factors, like auditors' expertise, complexity of the input crate(s), or their intended threat model.
\tool provides some objective guarantees that may lead to more converging audit results, as it
prevents missing relevant code, tracks effects across function calls and crate boundaries, and collects audit
results in a composable audit file.
However, it cannot further assist users with their final judgement and decision about the safety of each effect.

\begin{table}[t]
  \centering
  \caption{Audited functions and LoC between
  default (\S\ref{sec:efficient}) and full (\S\ref{sec:eval_cc}) audits.}
  \begin{tabular}[width=\textwidth]{
    >{\raggedright\arraybackslash}m{10em}
    >{\raggedright\arraybackslash}m{3em}
    >{\centering\arraybackslash}m{3em}
    >{\centering\arraybackslash}m{3em}
    >{\raggedleft\arraybackslash}m{3em}
  }
  \toprule
  \multicolumn{1}{c}{\multirow{2}{*}{\textbf{Crate}}} &
  \multicolumn{2}{c}{\textbf{Default Audits}} &
  \multicolumn{2}{c}{\textbf{Full Audits}} \\
  \cmidrule(lr){2-3} \cmidrule(lr){4-5}
  & Fns & LoC & Fns & LoC \\
  \midrule \midrule
    hyper-1.7.0 & 30 & 1453 & 54 & 1816\\
    http-1.3.1  & 9 & 258 & 27 & 583 \\
    fnv-1.0.7 & 0 & 0 & 0 & 0\\
    http-body-1.0.1 & 0 & 0 & 2 & 14 \\
    bytes-1.10.1 & 68 & 1425 & 114 & 1858\\
    itoa-1.0.15 & 1 & 9 & 1 & 9\\
    pin-project-lite-0.2.16 & 0 & 0 & 2 & 12 \\
  \hline
  \smallskip
  \end{tabular}

  \label{tab:audits_comparison}
\end{table}

%% file: discussion.tex
\section{Discussion and Future Work}

\tool{} helps developers audit Rust code more efficiently. It does so by identifying
potentially dangerous side effects through static analysis and tracking
context-dependent safety across the call graph.
While we found \tool{} effective in practice, there are some interesting
directions for future improvements.
Currently, \tool{} does not help developers audit ``deltas'', \ie changes between
different versions of a crate.
Different releases of the same crate usually share great similarities, and extending the tool
to focus only on code locations directly or indirectly affected by a change
could further reduce the auditing burden.
Finally, another direction is to analyze build-time effects. For example, Rust crates can execute
custom build scripts (\texttt{build.rs}) that sometimes generate potentially dangerous code.

%% file: rw.tex
\section{Related work}
\label{sec:rw}

\para{Empirical analysis and testing of the Rust ecosystem}
Rust packages are centrally organized by \texttt{crates.io}, with substantial
research analyzing \rustinline{unsafe} code in this ecosystem.
Prior work has analyzed \rustcode{unsafe} blocks usage~\cite{evans2020rust,astrauskas2020programmers},
and resulting bugs~\cite{qin2020understanding}.
In contrast to~\cite{evans2020rust}, we also treat closure and function pointer creation
as effects, construct a complete crate-level call graph for context-sensitive effects
propagation, and capture dangerous behavior beyond unsafe code alone.
Similarly, while~\cite{astrauskas2020programmers} resembles our effect model and categorizes
unsafe code using queries on the compiler's MIR representation, their analysis is limited to
unsafe operations.
Others research targets specific vulnerabilities, like Rust CVEs~\cite{xu2021memory},
yanked crates~\cite{li2022empirical}, and semantic version violations~\cite{semver-violations},
whereas we additionally analyze system effects in the ecosystem.
Dynamic testing tools, like RULF~\cite{jiang2021rulf} and SyRust~\cite{takashima2021syrust},
synthesize API call sequences for fuzzing specific bug classes (e.g., unreachable code and out of bounds checks),
complementing static analysis approaches and auditing.

\para{Effect analysis}
Prior work on side effect analysis---that is, determining how functions or expressions
impact a program's state beyond its return value, including memory, I/O, and control flow---focuses on languages like C/C++~\cite{rountev2001c,choi1993c,landi1993c} and Java
~\cite{jpure,rountev2004precise,le2005jit,milanova05parameterized,purity-analysis,YangHHK15,tao2012escape},
among others.
To account for mutation, many of these works
aim to compute sets of objects
that can be read and modified by each program statement;
this relies on computationally expensive points-to, alias, and escape analyses,
and produces verbose output.
These techniques are not necessary in Rust, where mutation and aliasing are delegated to the type system and are tractable outside of unsafe blocks.
Instead, we identify sets of effects at the function level, which in our experience is more useful for auditing.

\para{Static analysis for security audits}
There is prior work on static analyses designed specifically for security auditing,
for example, in the context of Java with tools like JoanAudit~\cite{JoanAudit}, Andromeda~\cite{andromeda}, FindBugs~\cite{findbugs} and others~\cite{spotbugs,livshits2005,taj,flowtwist,sflow}.
Most of this work implements taint and information-flow analysis based on preconfigured
sets of vulnerable patterns.
Although these tools aim to improve the efficiency and scalability of static analysis on Java
bytecode, they lack an interactive auditing model that guides users and allows for incremental and composable audit logs.

\para{Supply chain security}
Supply chain security spans a wide range of tools targeting different components.
In the Rust ecosystem, RustSec~\cite{rustsec} tracks vulnerabilities, while tools like
Cackle~\cite{cackle}, Cargo Audit~\cite{cargo-audit}, Cargo Crev~\cite{cargo-crev},
JFrog's Pyrsia~\cite{jfrogClosingSupply}, and Lib.rs~\cite{lib-rs} provide dependency management,
auditing metadata, and API usage checks.
In other ecosystems, particularly JavaScript, research focuses on publishing
and delivery~\cite{zimmermann2019small,cogo2019empirical,zahan2022weak}, though
dynamic features, like runtime dependency injection, hinder static analysis.
We target manual code auditing.

\tool's safety annotations can be compared with the ``auditing criteria''
used by Cargo Vet~\cite{cargo-vet};
in particular, they are most closely aligned with the built-in criteria
``safe-to-run'' and ``safe-to-deploy''.
We don't support Cargo Vet's other custom criteria, like ``crypto-reviewed'', targeted for
functional correctness,
but our effect model could be extended further, for example by treating
critical cryptographic functions as effects.
For Go, Google's Capslock~\cite{google-capslock,google-capslock-blog,google-capslock-talk}
performs a similar effect analysis.
However, unlike our approach, it is limited to a CLI interface without composable audit logs;
and it treats reflection as a runtime effect, while we statically resolve or overapproximate trait-based effects.

\para{Semantics and verification}
Much recent effort has gone into formalizing the
semantics of Rust's ownership mechanisms, including
RustBelt~\cite{rustbelt}, Oxide~\cite{oxide}, Stacked Borrows~\cite{stacked-borrows},
and the Miri interpreter for detecting undefined behavior~\cite{miri}.
Other projects like Rudra~\cite{bae2021rudra} analyze the interaction between unsafe Rust,
stack unwinding, and the Rust trait system.
These techniques can improve automation for detecting problems in unsafe Rust code,
but, like the existing ecosystem work, do not consider system side effects.
There is also work on verification tools for Rust,
including Smack~\cite{smack}, Kani~\cite{kani},
Prusti~\cite{prusti}, Aeneas~\cite{aeneas}, Verus~\cite{verus},
Hacspec~\cite{hacspec}, Flux~\cite{flux}, and RustHornBelt~\cite{rusthornbelt}.
Most verification tools only apply to safe Rust\dash---with the exception of~\cite{rusthornbelt}\dash---
as unsafe Rust is difficult to formalize.
Verification is labor-intensive and does not scale to the ecosystem,
but our techniques can help verifiers target code that is most critical.

%% file: acks.tex
\subsubsection*{Acknowledgments.}
We thank Bobby Holley for insightful discussions that sparked our effort, the
anonymous reviewers for their comments, and Pu Guo for his contribution to the
macro analysis.  This work was supported in part by the Berkeley Center for
Digital Assets, Mozilla, and the National Science Foundation (grant \#s 2048262,
2120642, 2155235, 2327336, and 2327338).

%% file: appendix.tex
\appendix

\renewcommand{\theHsection}{\Alph{section}}
\renewcommand{\theHsubsection}{\theHsection.\arabic{subsection}}

\section{Effect Categories}\label{sec:appendix}
\subsection{Unsafe and System Effects}\label{sec:effect_def}

\Cref{tab:effects} illustrates the complete set of effects that we support.
The \emph{unsafe effects} capture all possible operations outside
of Rust's safe subset, which are constrained to \rustcode{unsafe} blocks.
The \texttt{FFIDecl} effect indicates the declaration of a foreign function.
Variables in this function might not respect Rust's ownership rules if they
are used by non-Rust code.
This effect type is complementary to \texttt{FFICall}, for packages that only
publicly expose foreign functions, without calling them.
The \emph{system effects} originate from packages like the standard library or \texttt{libc},
and are treated as \texttt{SinkCall}s in our model.
System effects in Rust may internally make calls to the
\texttt{libc} library, which could be captured entirely
by the first effect category.
However, modeling them as a distinct category allows users
to audit their calling context instead of their implemention.
For example, when auditing file accesses users should consider
whether the files being accessed are appropriate within the
crate's context.
Thus, it could be impractical to surface the original unsafe effect
to the user, when auditing a file read or write operation for example,
as it does not provide meaningful information about the file
itself.
\texttt{ClosureCreation} indicates the existence of effects in closures,
which are executed after their creation.
Similarly, \texttt{FnPtrCreation} is used when function pointers might point
to functions containing other effects (see \S\ref{sec:model_higher_order}).

\subsection{Higher-Order Control Flow}\label{sec:model_higher_order}

Higher-order functions and polymorphic features are ubiquitous in both
the standard library and Rust application code.
Higher-order functions take function pointers or closures, which
themselves flow through the program as values, making it hard to determine
their safety without higher-order control flow analysis to determine \emph{all}
possible origins of those values.
Similarly, Rust's trait-based polymorphism, via generics or dynamic dispatch,
requires tracking all possible implementations to soundly analyze method calls.
\tool{} handles these two different forms of
higher-order control-flow -- (i)~function pointers and closures,
and (ii)~trait polymorphism -- via methods tailored to each.

\para{Closures and function pointers} In the following example, the closure at
line \ref{line:danger_closure_creation} takes a file path as an argument and reads
from that file.

\inputrust[frame=lines, basicstyle=\ttfamily\scriptsize, numbers=left, xleftmargin=\parindent]{code/closure_example.rs}

In this instance, the particular file being read depends on the argument to the
public function \rustcode{read_file}.
Because \rustcode{read_file} is a public function, any package calling
this library can use it to read any file on the system, and the onus is on
callers to make sure paths are properly sanitized.
If the function that returns the closure were instead written
\inputrust[frame=lines, basicstyle=\ttfamily\scriptsize, numbers=left, xleftmargin=\parindent]{code/closure_example_safe.rs}
\noindent then we could see the closure itself will properly sanitize any input
paths, and thus only access allowed files.

Rather than mark every possible call-site of an effectful function pointer or
closure as an effect itself, we instead mark their creation as effects (with
\texttt{ClosureCreation} and \texttt{FnPtrCreation}).
These functions are only created once, but may be called several times, so we
observe that, in practice, this approach results in less work for an auditor.

Moreover, closures are commonly used with iterators to process collections and are usually
pure, so naively marking them all as effectful will unnecessarily increase the audit
locations.
To avoid this, \tool{} scans the body of a closure and marks its creation as an effect
only if it contains operations that can be maliciously exploited.

\para{Traits} When trait method calls are resolved statically,
\tool{} will scan the specific implementation for potential effects.

\inputrust[frame=lines, numbers=left, escapeinside=||, 
xleftmargin=\parindent, basicstyle=\ttfamily\scriptsize]{code/traits_example.rs}

In our example, imagine a trait \rustcode{Trait} with a method \rustcode{meth}
and \rustcode{Struct} implementing that trait.
Since the call to \rustcode{meth} (line \ref{line:static_dispatch})
is statically dispatched, \tool{} will inspect \rustcode{Struct}'s
method implementation.
However, for generic or dynamically dispatched calls, \tool cannot
resolve the concrete type during analysis (e.g line \ref{line:dynamic_dispatch}).
Instead, it discovers \emph{all} available implementations of the trait in the
package and associates their effects with the abstract trait method.
The user then decides the method's safety based on the combined effects.